\documentclass[%
 aip,
 jmp,%
 amsmath,amssymb,
reprint,%
]{revtex4-1}

\usepackage{graphicx}
\usepackage{dcolumn}
\usepackage{bm}
\usepackage[pdftex,colorlinks=true,bookmarks=false,citecolor=blue,urlcolor=blue]{hyperref}

\begin{document}


\title{Selective engineering of cavity resonance for frequency matching in optical parametric processes}

\author{Xiyuan Lu}
\affiliation{
Department of Physics and Astronomy, University of Rochester, Rochester, NY 14627, USA
}%

\author{Steven Rogers}
\affiliation{
Department of Physics and Astronomy, University of Rochester, Rochester, NY 14627, USA
}%

\author{Wei C. Jiang}
\affiliation{
Institute of Optics, University of Rochester, Rochester, NY 14627, USA
}%

\author{Qiang Lin}%
 \email{qiang.lin@rochester.edu.}
\affiliation{
Institute of Optics, University of Rochester, Rochester, NY 14627, USA
}%
\affiliation{
Department of Electrical and Computer Engineering, University of Rochester, Rochester, NY 14627, USA
}%


\begin{abstract}
We propose to selectively engineer a single cavity resonance to achieve frequency matching for optical parametric processes in high-Q microresonators. For this purpose, we demonstrate an approach, selective mode splitting (SMS), to precisely shift a targeted cavity resonance, while leaving other cavity modes intact. We apply SMS to achieve efficient parametric generation via four-wave mixing in high-Q silicon microresonators. The proposed approach is of great potential for broad applications in integrated nonlinear photonics.

\end{abstract}


\maketitle

Optical parametric processes, including four-wave mixing (FWM), second-/third-harmonic generation (SHG/THG), parametric down-conversion (PDC), etc., have broad applications ranging from photonic signal processing \cite{Fejer06,AgrawalBook07}, tunable coherent radiation \cite{Dunn99}, frequency metrology \cite{Cundiff03}, to quantum information processing \cite{Martini05}. In recent years, there has been increasing interest in optical parametric processes in high-Q micro-/nano-cavities, which exhibit great potential for dramatically enhancing parametric generation \cite{Vahala03, Vahala04, Maleki04, Vahala07, Turner08, Leuchs10, Breunig11, Kippenberg11, Jiang12, Pertsch13, Vuckovic13, Noda14, Solomon14, Kumar14}. However, their efficiencies rely crucially on frequency matching among the interacting cavity modes, which is generally deteriorated by device dispersion. Frequency matching is particularly challenging in high-Q cavities, due to the narrow linewidths of cavity resonances.

To date, a variety of methods have been proposed for frequency matching. For FWM, a $\chi^{(3)}$ nonlinear process, current methods primarily focus on engineering group-velocity dispersion of the devices \cite{Vahala04, Maleki04, Turner08, Kippenberg11, Jiang12, Jiang13}. For SHG, a $\chi^{(2)}$ nonlinear process, intermodal dispersion or birefringence is generally employed to mitigate the frequency mismatch \cite{Leuchs10, Breunig11, Pertsch13, Vuckovic13, Noda14, Solomon14, Kumar14}. These methods rely on manipulating material/waveguide dispersion of devices, which collectively shifts the resonance frequencies of all cavity modes by different extents.

\begin{figure}[b]
\includegraphics[width=1\columnwidth]{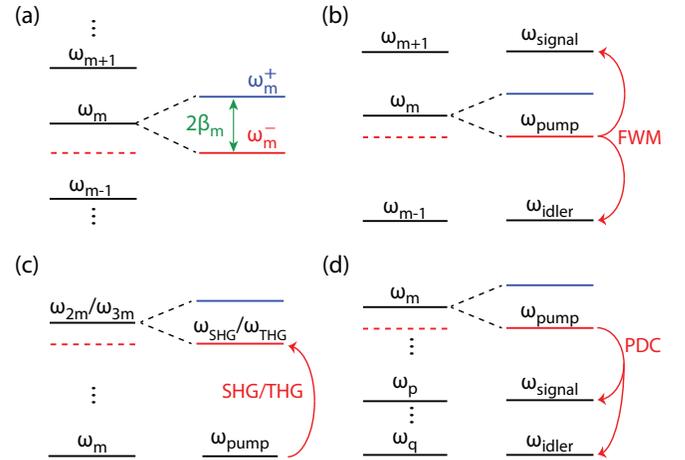}
\caption{\label{Fig1} (a) Schematic illustration of selective mode splitting. The frequency levels represent cavity resonances with different azimuthal mode numbers ($m$-1, $m$, $m$+1). The $m^{th}$ cavity mode is split into two modes. The magnitude of splitting is precisely controlled so that the lower frequency mode matches the desired frequency (red dotted line). (b) Application of SMS for frequency matching in degenerate FWM. The $m$+1 and $m$-1 modes are used as the signal and idler modes, respectively. The red dotted line represents the desired frequency, $(\omega_{m+1}+\omega_{m-1})/2$, for the pump mode. The arrows in red indicate the process of parametric generation. (c, d) Application of SMS for frequency matching in SHG/THG and PDC, respectively. }
\end{figure}

In this paper, we propose and demonstrate an effective approach for frequency matching that can be applied universally to optical parametric processes. Instead of modifying all cavity resonances via dispersion engineering, we directly shift a single cavity resonance to a desired frequency, while leaving other cavity modes intact. This is realized by splitting the targeted cavity mode at $\omega_m$ into two modes at new frequencies $\omega_m \pm \beta_m$. By controlling the magnitude of splitting, one resonance can be shifted to coincide with the desired frequency (Fig.~\ref{Fig1}(a)). We term this approach selective mode splitting (SMS).

SMS can be applied to various $\rm \chi^{(2)}$/$\chi^{(3)}$ processes (Fig.~\ref{Fig1}(b-d)). For example, degenerate FWM requires three interacting cavity modes to be equally spaced in frequency, $\rm \omega_{signal}-\omega_{pump}=\omega_{pump}-\omega_{idler}$, which is often not satisfied due to device dispersion. The problem can be solved by shifting one cavity mode to the desired frequency, thus enabling parametric generation (Fig.~\ref{Fig1}(b)). SMS is particularly useful for nonlinear processes including SHG/THG and PDC (Fig.~\ref{Fig1}(c)(d)), where frequency mismatch is difficult to mitigate due to the large frequency separation among interacting modes.

\begin{figure}[b]
\includegraphics[width=1\columnwidth]{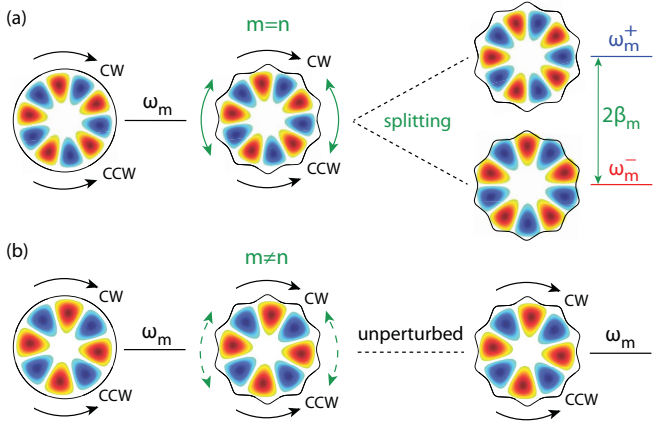}
\caption{\label{Fig2} Schematic illustration of sinusoidal radius modulation, $\delta r = \alpha \cos (2 n \phi)$, to achieve SMS. (a) The radius modulation couples two degenerate counter-propagating cavity modes, CW and CCW, with azimuthal mode number $m$ equal to $n$, which produces two standing-wave modes with a mode splitting of $2 \beta_m$. (b) Other cavity modes ($m \neq n$) remain unperturbed. }
\end{figure}

In practise, SMS can be realized by the approach illustrated in Fig.~\ref{Fig2}. A rotationally symmetric microresonator exhibits two degenerate cavity modes for each resonance frequency $\omega_m$, with the same azimuthal number $m$. These two modes propagate in opposite directions, clockwise (CW) and counter-clockwise (CCW), respectively. We propose to apply a sinusoidal modulation to the device radius, $r = r_0 + \alpha \cos (2 n \phi)$. Such a radius modulation would introduce a coupling between the CW and CCW modes. Following the perturbation theory with shifting boundaries \cite{Johnson02}, the mode coupling strength $\beta_m$ is found to be
\begin{eqnarray}
\beta_{m} = g \alpha \delta_{mn}, \label{eq3}
\end{eqnarray}
where $g$ is given by
\begin{eqnarray}
g = \pi \omega_m \frac{\int{d A \left[ \Delta\epsilon(|E_{z}|^2+|E_{\phi}|^2) - \Delta(\epsilon^{-1})|D_{r}|^2\right]}}{\int {d V \epsilon (|E_{z}|^2+|E_{\phi}|^2+|E_{r}|^2) } }. \label{eq4}
\end{eqnarray}
In Eq.~(\ref{eq4}), $\rm \Delta\epsilon \equiv \epsilon_{d}-\epsilon_{\rm c}$ and $\rm \Delta(\epsilon^{-1}) \equiv \epsilon_{d}^{-1}-\epsilon_{c}^{-1}$, where $\epsilon_{\rm d}$ and $\epsilon_{\rm c}$ represent the dielectric constants for the device and cladding layers, respectively. $E_r$, $E_\phi$, and $E_z$ are the electric field components along the radial, azimuthal, and vertical direction, respectively. $D_r$ is the displacement vector component along the radial direction. The integrations with respect to $dA$ and $dV$ are over the area of the shifting boundary and the entire volume, respectively.

This mode coupling lifts the frequency degeneracy of the two modes and renormalizes them into two standing-wave modes with new frequencies of $\omega_m^{\pm} = \omega_m \pm \beta_m$ (Fig.~\ref{Fig2}(a)). Physically, one of the standing-wave modes experiences a smaller effective radius and thus has a higher resonance frequency $\omega_m^+$, and vice versa. Equation~(\ref{eq3}) shows that, when $m=n$, $\beta_m$ scales linearly with $\alpha$ and $g$. The finite-element-method simulation shows that $g$ is as large as ${\rm 10.3~GHz/nm}$ for a silicon microdisk with a radius of $\rm 9.4~\mu m$, thus enabling efficient tuning of the resonance frequency. Of particular interest is that the sinusoidal radius modulation leaves all other cavity modes intact ($\rm \beta_{m} = 0$ for $m \ne n$), as illustrated in Fig.~\ref{Fig2}(b). This single-resonance selection feature distinguishes our approach from other mode-splitting methods \cite{Haroche95, Painter05, Painter07, Adibi12, Qiu08, Su08, Yang10, Goddard11} that were studied in the contexts of defect scattering, signal filtering, and particle sensing.

\begin{figure}[t]
\includegraphics[width=1\columnwidth]{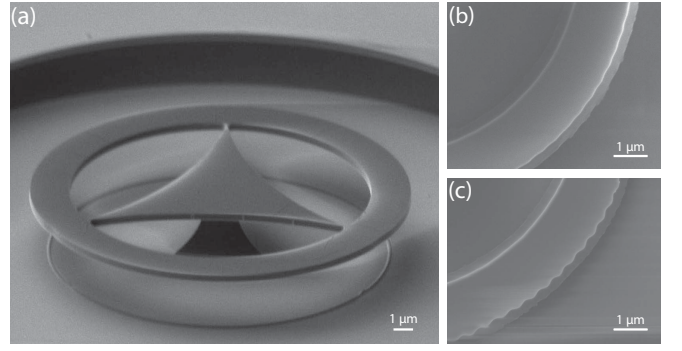}
\caption{\label{Fig3} (a) Scanning electron microscopic (SEM) image of a silicon microring sitting on a silica pedestal. (b, c) SEM images of microrings with outer boundaries sinusoidally modulated with amplitudes of 5~nm and 40~nm, respectively. }
\end{figure}

\begin{figure}[t]
\includegraphics[width=1\columnwidth]{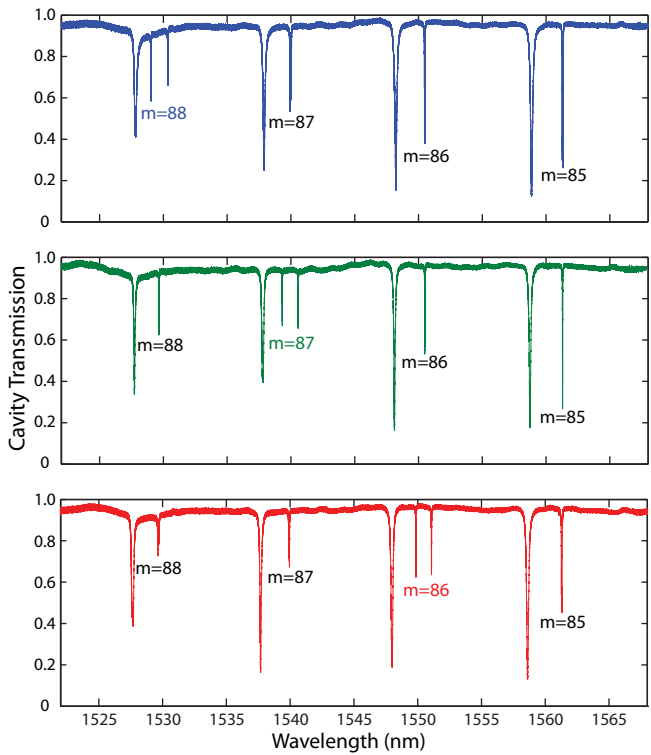}
\caption{\label{Fig4} Cavity transmission of three devices with different periods of radius modulation. Two mode families are clearly visible. The mode family of interest is labeled with azimuthal numbers. The targeted modes, with mode splittings above 1~nm, have azimuthal numbers of 88, 87, and 86, respectively, from top to bottom. }
\end{figure}

To prove this concept, we fabricated a set of silicon microring resonators with outer radii modulated by different periods and amplitudes. Figure~\ref{Fig3}(a) shows an example of a 220-nm-thick device with an outer radius of $\rm 9.4~ \mu m$ and a ring width of $\rm 2~\mu m$. The spoked-ring geometry is used to eliminate the high radial-order optical mode families, for a clear view of the selected mode splitting. Figure~\ref{Fig3}(b,c) show the details of the sinusoidally modulated outer boundaries. The fabricated devices are tested in a tapered fiber coupling setup \cite{Jiang12}. Figure~\ref{Fig4} shows the cavity transmission spectra of three devices with different modulation periods. By changing the modulation periods, we are able to selectively split the cavity modes located around 1530, 1540, and 1551~nm, respectively, with mode splittings around 1.2~nm. In particular, other cavity modes with different azimuthal numbers remain nearly intact.

\begin{figure}[t]
\includegraphics[width=1\columnwidth]{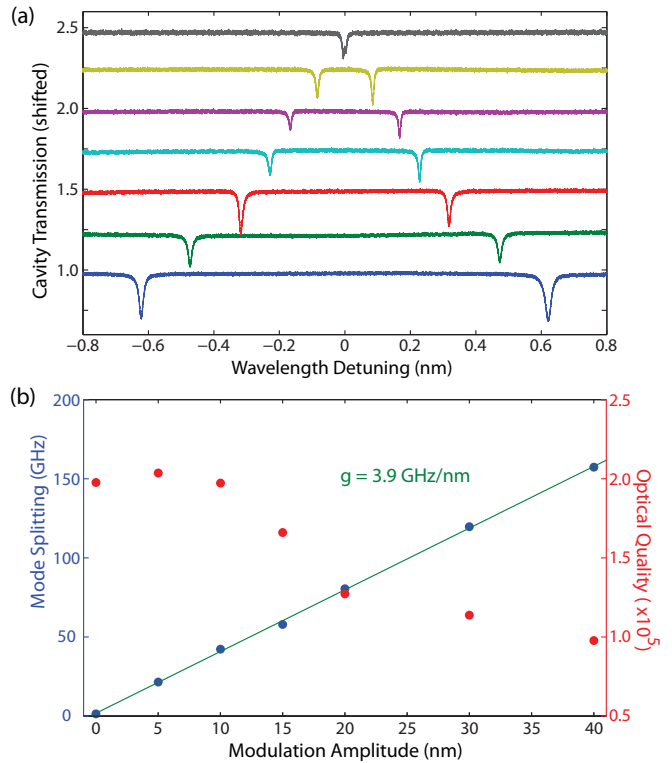}
\caption{\label{Fig5} (a) Cavity transmission spectra of a targeted optical mode ($m=87$) with different modulation amplitudes varying from 0 to 40 nm. Transmission spectra are shifted vertically for better comparison of traces. (b) Mode splitting as a function of radius modulation amplitude. Experimental data are shown as blue dots with a linear fitting shown in green. The corresponding optical quality factors are plotted in red. }
\end{figure}

Figure~\ref{Fig5}(a) shows the transmission spectra of the cavity mode around 1540~nm ($m = 87$)  with modulation amplitude ($\alpha$) varying from 0 to 40~nm. Without radius modulation, the cavity mode exhibits a small natural splitting, which stems from scattering induced by the sidewall roughness of the device \cite{Painter05}. The mode splitting increases linearly with the modulation amplitude, reaching a value of 1.24~nm at a modulation amplitude of 40~nm, which corresponds to a resonance tuning rate of 3.9~${\rm GHz/nm}$. This tuning rate is smaller than the theoretical prediction of 10.3~${\rm GHz/nm}$, which is likely due to the etching process that tends to smooth out the sinusoidal patterns. On the other hand, the optical Q maintains around $2\times 10^5$ for modulation amplitudes up to 15~nm (Fig.~\ref{Fig5}(b)). It slightly decreases with a larger modulation amplitude but still maintains as high as $\sim 1\times10^5$ for a modulation amplitude up to 40~nm. The significant resonance tuning range and well-maintained optical Q render SMS a promising technique for optical parametric processes.

\begin{figure}[t]
\includegraphics[width=1\columnwidth]{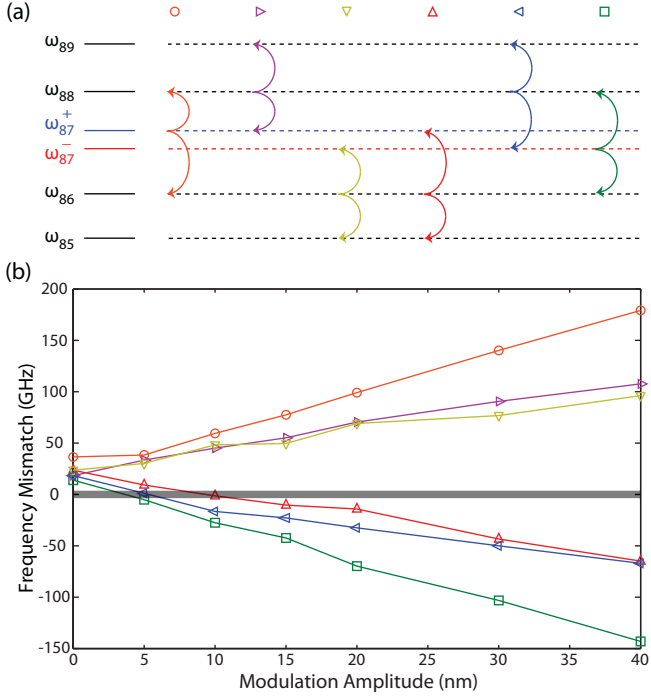}
\caption{\label{Fig6} (a) Six cavity mode combinations for degenerate FWM. (b) Frequency mismatch versus radius modulation amplitude for the six schemes. The gray area indicates where the frequency mismatch falls within the cavity linewidths ($\sim$~3~GHz). The lines are used for eye guidance only. }
\end{figure}

Moreover, SMS allows a variety of cavity mode combinations for the parametric processes, enabling flexible frequency management. For example, in degenerate FWM, there are six potential mode combinations within five adjacent cavity modes (Fig.~\ref{Fig6}(a)). The corresponding frequency mismatches versus different modulation amplitudes are shown in Fig.~\ref{Fig6}(b). Without radius modulation, the device exhibits a frequency mismatch around 20~GHz, prohibiting the FWM process. However, modulation amplitudes of 5 and 10 nm are able to achieve a frequency mismatch of 1.1~GHz and -1.4~GHz, respectively, for two different mode combinations (blue and red), both smaller than the cavity linewidth.

\begin{figure}[t]
\includegraphics[width=1\columnwidth]{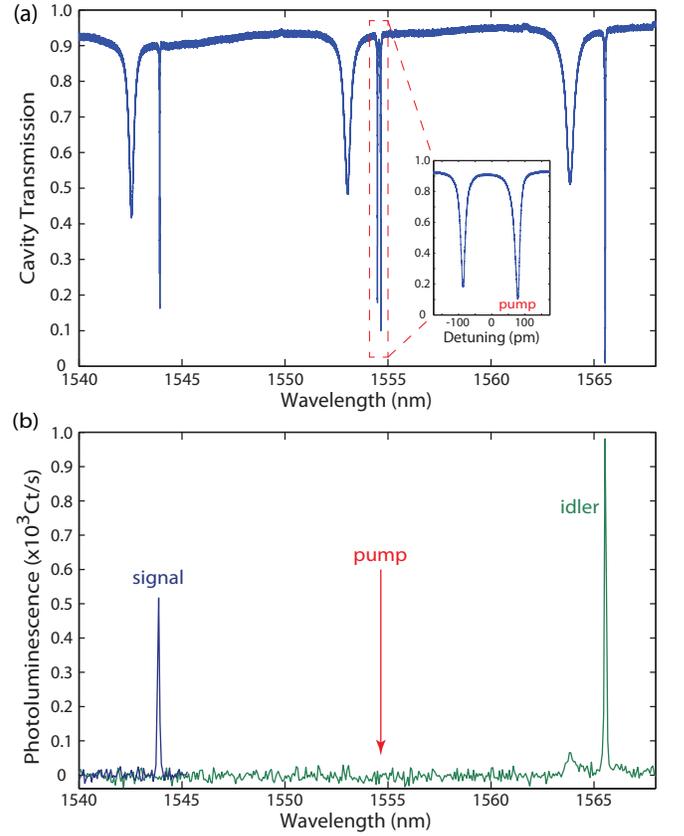}
\caption{\label{Fig7} (a) Cavity transmission of the device with a radius modulated by $2 \times 86$ periods with an amplitude of 5~nm. The inset shows the details of the splitting, where the mode on the right is used as the pump for degenerate FWM. (b) Photoluminescence spectrum of the signal and idler photon pairs produced via spontaneous FWM.}
\end{figure}

The achieved frequency matching in high-Q cavities guarantees efficient FWM processes that are otherwise forbidden. A typical example is shown in Fig.~\ref{Fig7}, where the device radius is modulated by $\rm 2 \times 86$ periods with an amplitude of 5~nm. The cavity mode located around 1552.7~nm is split into two modes separated by approximately 150~pm (Fig.~\ref{Fig7}a, inset). The mode splitting reduces the frequency mismatch significantly from 21.5~GHz to 1.48~GHz if the right split mode is employed. Consequently, by pumping at this mode (Fig.~\ref{Fig7}a, inset), we are able to produce efficient parametric generation of bright signal and idler photon pairs at wavelengths of 1541.9 and 1563.8~nm, respectively (Fig.~\ref{Fig7}(b)). The smaller intensity of the signal mode compared with the idler is simply due to the different coupling rates of photons to the delivery fiber \cite{Jiang12}. In contrast, when the left split mode is used, the frequency mismatch is about 41.5~GHz and FWM is quenched.

Although our discussion focuses on engineering a single cavity mode, in principle, SMS can be applied to selectively split multiple cavity modes, by using a radius modulation of $r=r_0 + \alpha_1 \cos(2m_1 \phi) + \alpha_2 \cos(2m_2 \phi) + \cdots$. Such a multi-mode SMS can be used to engineer all the interacting cavity modes simultaneously in an optical parametric process. The resulting modes are standing waves and therefore have a better mode overlap, which would further increase the efficiency of parametric generation. On the other hand, the inner radius of a microring can also be employed for modulation. As a whispering gallery mode is mostly confined by the outer perimeter, the inner modulation would produce a much smaller resonance perturbation, resulting in a finer control of cavity resonances compared with the outer modulation. Therefore, the inner and outer modulation, when combined together, enable frequency tuning with both high resolution and large dynamic range.

In summary, we have demonstrated selective mode splitting (SMS) as a universal approach for frequency matching of nonlinear optical parametric processes in high-Q microresonators. It is capable of flexibly engineering individual cavity resonances to compensate a large frequency mismatch. We applied this approach to produce bright photon pairs via degenerate four-wave mixing. SMS exhibits great potential not only for broad applications in integrated nonlinear photonics, but also for many applications that requires precise frequency control of cavity resonances, such as cavity quantum electrodynamics with embedded atoms/quantum dots/spin defects \cite{Painter07, Lu14}.

This work was supported by National Science Foundation under grant ECCS-1351697. It was performed in part at the Cornell NanoScale Science \& Technology Facility (CNF), a member of the National Nanotechnology Infrastructure Network, which is supported by the National Science Foundation.

\end{document}